\documentclass[11pt]{article}
\usepackage{amssymb,amsmath}

\def\appendix#1{\addtocounter{section}{1}\setcounter{equation}{0}
\renewcommand{\theequation}{\thesection.\arabic{equation}}

\renewcommand{\thesection}{\Alph{section}}
\section*{
\thesection\protect\indent \parbox[t]{11.715cm} {#1}}
\addcontentsline{toc}{section}{Appendix\thesection\ \ \ #1} }

\def\nn{\nonumber}
\newcommand{\Tr}[1]{\:{\rm Tr}\,#1}
\newcommand{\eqn}[1]{(\ref{#1})}
\def\beqa{\begin{eqnarray}}
\def\eeqa{\end{eqnarray}}
\def\nn{\nonumber}
\def\be{\begin{equation}}
\def\ee{\end{equation}}
\newcommand{\half}{{\textstyle\frac{1}{2}}}
\newcommand{\real}{{\mathbb{R}}} 
\newcommand{\complex}{{\mathbb{C}}} 
\newcommand{\del}{\partial}
\begin{document}

\begin{center} {\Large \bf  Star products, duality and double Lie algebras
}
\end{center}

\bigskip

\begin{center} {\bf Olga V. Man'ko$^1$, Vladimir I.~Man'ko$^1$, Giuseppe
Marmo$^2$ and Patrizia Vitale$^2$}
\end{center}
\medskip

\begin{center} $^1$ {\it P. N. Lebedev Physical Institute\\ Leninskii Prospect 53, Moscow
119991, Russia\\} {\tt omanko@sci.lebedev.ru, manko@sci.lebedev.ru\\
} $^2$-{\it Dipartimento di Scienze Fisiche, Universit\`{a} di
Napoli {\sl Federico II}\\ and {\it INFN, Sezione di Napoli}\\
Monte S.~Angelo, Via Cintia, 80126 Napoli, Italy}\\ {\tt
giuseppe.marmo@na.infn.it, patrizia.vitale@na.infn.it}
\end{center}

\begin{abstract}
Quantization of classical systems using the star-product of symbols
of observables is discussed. In the star-product scheme an analysis
of dual structures is performed and a physical interpretation is
proposed. At the Lie algebra level duality is shown to be connected
to double Lie algebras. The analysis is specified to quantum
tomography. The classical tomographic Poisson bracket is found.
\end{abstract}

\medskip


\section{Introduction}
The transition from quantum to classical mechanics has been an
important research subject since the beginning of quantum
mechanics (see \cite{Beppebook,MalManbook} for a review). A
suitable setting for this problem is represented by the
Wigner-Weyl-Moyal formalism where the operators corresponding to
observables and the states, considered as linear functionals on
the space of observables, are mapped onto functions on a suitable
manifold. Such a representation for quantum mechanics has been
later generalized yielding to the deformation quantization program
\cite{Fronsdal}. There the operator noncommutativity is
implemented by a noncommutative (star) product which is a
generalization of the Moyal product  \cite{Groen,Moyal,Zackos}.
Since then, most attention to the star-product quantization scheme
has been devoted to the case where the functions (symbols of the
operators) are defined on the ``classical" phase space of the
system \cite{Curtr,Beresin,Voros,DodMan}. But this is not the only
possibility. An interesting example which cannot be described in
terms of a deformation of the commutative product of classical
phase space is the so called quantum tomography. In this setting
quantum states are mapped onto probability distributions which
depend on a random variable $X$ representing the position in
classical phase--space \cite{Mancini1} and two additional
 real parameters labelling different reference frames in
the phase space. This approach provides a formulation of quantum
mechanics where the states are described by positive probabilities
as an alternative to wave functions and density states
\cite{Mancini2,Mancini3}. In
\cite{OlgaBeppe1,OlgaBeppe2,OlgaBeppe3} it has been shown that the
symplectic tomography as well as other known types of tomographic
methods for measuring quantum states like optical tomography
\cite{Bertrand,VogelRisken} and spin tomography
\cite{DodonovPhysLet,OlgaJETP,Paini} can be formulated in a
star-product approach once the star-product quantization scheme is
suitably reinterpreted.  This new presentation led recently
\cite{Patr} to find and elucidate a specific duality symmetry of
the star-product quantization which has proved to be useful to
clarify well known relations among important distribution
functions of quantum mechanics such as  the Wigner function
\cite{Wig32}, the Husimi $Q$-function \cite{Hus40}, the
Sudarshan--Glauber $P$-function \cite{Sud63,Gla63} and s-ordered
quasidistributions \cite{GlaCah69}, as well as relations among
time--frequency quasidistributions known in signal analysis
\cite{Cohen,MendesJPhysA}. Also the duality symmetry provides a
tool to  find new  solutions to the  nonlinear associativity
equation for the star-product kernel if one solution of the
equation is available.

The formulation of the  star-product quantization scheme proposed in
\cite{OlgaBeppe1,OlgaBeppe2,Patr} is based on the properties of two
sets of operators which we refer as {\it quantizers} and {\it
dequantizers}, respectively. They are in fact   operator-valued
functions on a manifold and behave like elements of dual vector
spaces. In the Weyl--Moyal star-product quantization the quantizer
and the dequantizer coincide up to a constant (implying that the
Moyal product is self-dual) and the manifold is the phase-space of
the system. Moreover,  in the limit $\hbar\rightarrow 0$ the
commutator correctly reproduces the canonical Poisson bracket for
the classical phase space. This should be possible for other
quantization schemes as well; namely, knowing the dependence of the
star product kernel on the deformation parameters one should recover
the classical Poisson bracket in the appropriate limit.  Till now
for the symplectic tomographic approach \cite{Olga97,Mendes} the
classical Poisson structure was not obtained in explicit form.

In this paper we focus our analysis on the duality relation
between quantizers and dequantizers. As we shall see, this allows
to introduce dual Lie algebra structures on the symbols, which we
will investigate in some detail.  Moreover, we take the occasion
to clarify the physical meaning of dual {\it quantization} schemes
and specifically of the dual tomographic scheme, as the one
encoding the information about   quantum observables, as opposed
to tomograms which describe  quantum states. We then analyze the
classical limit in the tomographic representation. We obtain the
Poisson structure for classical symplectic tomograms associated to
classical observables as a suitable limit of the quantum
tomography. As a related subject we discuss the nonlinearity
property of the dual tomographic star-product kernel with respect
to the quantizer operator. This influences the uniqueness of the
corresponding Poisson structure.

The paper is organized as follows.

In Section 2 we review the star-product quantization scheme
following \cite{OlgaBeppe1,OlgaBeppe2}, together with a class of
star products ($K$-star products) which is obtained via a specific
deformation procedure \cite{OlgaBeppe2}. In Section 3 we deepen the
analysis of the duality symmetry  started in \cite{Patr} and work
out explicit examples. In section 4, starting from the observation
that the anti-symmetrized star product kernel, for a discrete space
of parameters,   may be interpreted as the structure constants of a
Lie algebra we discuss a class of solutions of the Jacobi identity.
In section 5, upon reviewing the symplectic tomography scheme, we
analyze  the classical limit of the tomographic quantum Poisson
brackets. Then we investigate the form of the integral kernel of the
Moyal product in Fourier representation. As in the tomographic case,
we find that the star product kernel differs from the point-wise one
by a twist factor which is the exponential of the symplectic area.
Finally  we summarize our results.

\section{The star product formalism}

Following~\cite{OlgaBeppe1,OlgaBeppe2,OlgaBeppe3,Patr} let us
consider  a given manifold $M$, with coordinates $\vec x=(
x_1,x_2,\ldots,x_N)$ and two dual sets of operators $\hat D (\vec
x)$ and $\hat U(\vec x)$, which generate, as a continuous basis,
the vector space  $V$ and its dual $V^\ast$. The coordinates may
be discrete as well. These are vector spaces of operators acting
on a given Hilbert space. We call the operators $\hat D(\vec x)$
{\it quantizers} and the operators $\hat U(\vec x)$ {\it
dequantizers} due to the following reasons. On $V^* \oplus V$, by
exploiting the duality relation, it is possible to  define a
scalar product $<~|~>$
\beqa
<(a,0)|(0,b>&=& a(b) \label {prod} \nn\\
<(a,0)|(c,0)>&=& <(0,b)|(0,c)>= 0 \label{product}
\eeqa
where $(a,b)\in V^*\oplus V$. Being vector spaces of linear maps,
they both carry  an associative product and therefore a  Lie
algebra structure. Thus, $V$ and $V^*$ are assumed to both carry a
Lie algebra structure coming from the associative product.  By
requiring that the product \eqn{product} be
 adjoint invariant we may define a Lie algebra structure on
$V^*\oplus V$. Therefore our considerations are closely related to
what are known as metrical Lie algebras.   Having  a pairing
between $V$ and $V^\ast$ we can associate with an operator $\hat
A$ in $V$ a function $f_A(\vec x)$ in the following way. The
operator $\hat A\equiv (0, \hat A)$ may be written in terms of the
basis in $V$ as
\be
\hat A= \int d\vec x~ a(\vec x) \hat D (\vec x) \label{eq.2}
\ee
with
\be
a(\vec x) \equiv f_A(\vec x) = <(\hat U(\vec x),0)|(0,\hat A)>
\ee
which may be rewritten as
\be
f_A(\vec x)= \Tr \hat U^\dag (\vec x) \hat A \label{eq.1}
\ee
and the trace stays for the appropriate scalar product. Replacing
\eqn{eq.2} into \eqn{eq.1} we get a compatibility condition
between $\hat U(\vec x)$ and $\hat D(\vec x)$ which is just the
duality relation between the two bases of $V$ and $V^*$
\be
<(\hat U(\vec x),0)|(0,\hat D(\vec x')>=\Tr \hat U^\dag(\vec
x)\hat D(\vec x')= \delta(\vec x-\vec x') \label{eq.3}
\ee
where  $\delta(\vec x-\vec x^\prime)$ is a Dirac delta-function
for the continuous variables,  a Kronecker delta
 for discrete variables.
  So the operators $\hat
U(\vec x)$ associate, through the scalar product,   to  the
operator $\hat A$ a function, i.e. they {\it dequantize} the
quantum observable, while  the role of the operators $\hat D(\vec
x)$ is opposite. They furnish a basis for the expansion of $\hat
A$ with coefficients $f_A(\vec x)$ ($\vec x$ plays the role of a
continuous index. This means that  they  associate to  the
function $f_A(\vec x)$ (classical observable) the operator $\hat
A$ (quantum observable), i.e. they ``quantize" the function. This
association should remind us of the momentum map associated with a
Lie algebra $ \mathcal{G}$ acting on the cotangent bundle $T^\ast
\mathcal{G}=\mathcal{G}^*\oplus \mathcal{G}$. On the corresponding
functions we define a $\ast$-product instead of a point-wise
product.

 The associative
star-product of symbols $f_A(\vec x), f_B(\vec x)$ associated with
 the operators $\hat  A$ and $\hat  B$ is defined as
\begin{equation}
\hat A\hat B\leftrightarrow f_A(\vec x)\ast f_B(\vec
x):=f_{AB}(\vec x).
\end{equation}
  This product  is associative since the
product of the operators is associative, i.e.,
\begin{equation}\label{5}
(\hat A\hat B)\hat C=\hat A(\hat B\hat C)\rightarrow(f_A(\vec
x)\ast f_B(\vec x))\ast f_C(\vec x)=f_A(\vec x)\ast(f_B(\vec
x)\ast f_C(\vec x)).
\end{equation}
 Moreover it is nonlocal and it can be
described through an  integral kernel
\begin{equation}\label{eq.6}
f_A(\vec x)\ast f_B(\vec x)=\int_{M\times M} K(\vec x_1,\vec
x_2,\vec x)f_A(\vec x_1)f_B(\vec x_2)\, d \vec x_1\, d \vec x_2
\end{equation}
which plays the role of the structure function for the product.
 Using Eqs. (\ref{eq.1}),(\ref{eq.2}) we can get the
expression for the kernel in the form
\begin{equation}\label{eq.7}
K(\vec x_1,\vec x_2,\vec x)= <(\hat U(\vec x),0)| (0,\hat D(\vec
x_1) \hat D(\vec x_2))> =  \mbox{Tr}\Big(\hat U^\dag(\vec x) \hat
D(\vec x_1)\hat D(\vec x_2) \Big).
\end{equation}
 We can see that this expression  is linear with respect to $\hat U(\vec x)$ and
quadratic with respect to $\hat D(\vec x)$. From this point of
view there is an asymmetry in  the kernel with respect to
quantizers and dequantizers.\footnote{ We warn the reader that in
 our previous articles on this subject, with a less precise notation,
 we used the symbol
$\hat U(x)$ instead than its adjoint.} The associativity condition
for operator symbols implies that the kernel $K(\vec x_1,\vec
x_2,\vec x )$ satisfies the nonlinear equation
\begin{equation}\label{C2}
\int K(\vec x_1,\vec x_2, \vec y)K(\vec y,\vec x_3, \vec x_4)d\vec
y=\int K(\vec x_1,\vec y, \vec x_4)K(\vec x_2,\vec x_3, \vec
y)d\vec y.
\end{equation}

In connection with our previous comment on metrical Lie algebras
we might say that we are dealing here with metrical Lie algebras
realized via associative algebras of operators or associative
algebras of functions via the star product formalism.

Let us consider as an example the two real matrix algebras
\be
{a}= \left\{\begin{pmatrix}
  1 & 0 \\
  0 & 0
\end{pmatrix}~,~~~~
\begin{pmatrix}
  0 & 1 \\
  0 & 0
\end{pmatrix}\right\}\label{a}
\ee
and
\be
{b}=\left\{\begin{pmatrix}
  0 & 0 \\
  0 & 1
\end{pmatrix}~,~~~~
\begin{pmatrix}
  0 & 0\\
  1 & 0
\end{pmatrix}\right\}\label{b}
\ee
They commute with each other and are dual with respect to the
adjoint invariant product $\gamma$, defined on the direct sum
$a\oplus b$
\be
\gamma(A_i, B_j)=\Tr(A_i J B_j J) \label{gamma}
\ee
with $i=1,2$ labelling the bases \eqn{a}, \eqn{b} and
\be J=\Bigl(
\begin{array}{cc}
  0 & 1 \\
  1 & 0
\end{array}
\Bigr)
\ee
Their direct sum is the Lie algebra $\mathcal{GL}(2,\mathbb{R})$.
The pairing \eqn{gamma} allows to associate to a given operator
$B$ in the vector space $b$ a function, $f_B$, which represents
the symbol of $B$ in the given scheme. Consider indeed
\be
\hat B= x_1 B_1 + x_2 B_2
\ee
then
\beqa
f_B(1)&=&\Tr(A_1 J B J)=x_1\nn\\
f_B(2)&=&\Tr(A_2 J B J)=x_2
\eeqa
thus implying that
\be
\hat B= f_B(1) B_1 + f_B (2) B_2
\ee
\subsection{Deformed star-products from deformed operator products}
 In~\cite{OlgaBeppe2} a deformed operator product was introduced
 ($K$-product) in the form
\begin{equation}\label{D1}
\hat A\cdot_K \hat B=\hat A\hat K\hat B
\end{equation}
where $\hat K$ is a generic operator. It  satisfies the
associativity condition
\begin{equation}\label{D2}
\Big(\hat A\cdot_K\hat B\Big)\cdot_K\hat C=\hat A\cdot_K\Big(
B\cdot_K \hat C\Big).
\end{equation}
In matrix representation the deformed operator  product  provides
a deformed  matrix product of the form
\begin{equation}\label{D3}
A\cdot_K B:=A K B
\end{equation}
where $A, B, K$ are the  matrices corresponding to the operators
of Eq. ~(\ref{D1}). Using the $K$ product and considering the
operator
 symbols in a given star product scheme,  we can define a deformed star product
\begin{equation}
\Big(f_A\ast_K f_B\Big)(\vec x)=:\mbox{Tr}(\hat A\hat K\hat B\hat
U(\vec x)).
\end{equation}
Due to the associativity of the $K$-product of operators the
 $K$-star product of the symbols is obviously associative,
i.e.,
\begin{equation}\label{D4}
\Big(\Big(f_A\ast_K f_B\Big)\ast_K f_C\Big)(\vec x)=\Big(f_A\ast_K
\Big(f_B\ast_K f_C\Big)\Big)(\vec x)).
\end{equation}
As in the undeformed case we can derive an expression for the
kernel
\begin{equation}\label{D6}
\Big(f_A\ast_k f_B\Big)(\vec x)=\int K^{(K)}(\vec x_1,\vec x_2,
\vec x)f_A(\vec x_1)f_B(\vec x_2)d \vec x_1d\vec x_2
\end{equation}
with
\begin{equation}\label{D7}
K^{(K)}(\vec x_1,\vec x_2, \vec x)=\mbox{Tr}\Big(\hat D(\vec
x_1)\hat K\hat D(\vec x_2) \hat U(\vec x)\Big)
\end{equation}
It can be verified that it satisfies the associativity equation
\eqn{C2}.

 K-deformed products are a simple way to generate new
associative products which deserves in our opinion further
attention. They are a simple instance of deformed products
associated with a Nijenhuis tensor \cite{CGM}.   In section 5 we
give an example of such potentiality in the framework of metrical
Lie algebras.

\section{Duality symmetry of the star-product}
In~\cite{Patr}  a duality symmetry of the star-product scheme was
found. Shortly, it originates from the observation that  the role
of  the quantizer and the dequantizer can be exchanged  because
the compatibility condition~(\ref{eq.3}) remains fulfilled. Then,
 one can introduce a new pair of operators $\hat U^\prime(\vec x)
$ and $\hat D^\prime(\vec x) $
\begin{equation}\label{C1}
\hat U^\prime(\vec x)= \hat D(\vec x),\quad \hat D^\prime(\vec
x)=\hat U(\vec x).
\end{equation}
We say that  new pair  quantizer--dequantizer is dual to the
initial one. The terminology is justified by the observation that
\be
(V\times V^*)^* \simeq  V^*\times V^{**}\simeq V^*\times V \ee
when $V^{**}=V$.
 This interchange corresponds to a specific
symmetry of the equation for the associative star-product kernel.
We have seen that the kernel $K(\vec x_1,\vec x_2,\vec x )$
satisfies the associativity equation \eqn{C2}, a solution to which
is represented by (\ref{eq.7}). But this equation admits also the
dual solution
\begin{equation}\label{C3}
K^{(d)}(\vec x_1,\vec x_2,\vec x)=\mbox{Tr}\Big(\hat U(\vec
x_1)\hat U(\vec x_2)\hat D^\dag(\vec x)\Big).
\end{equation}

 Solutions may be singled out by requiring particular
symmetry properties; for example imposing translation invariance
one finds that the only solution is the Moyal one \cite{SZS71}.

The analogue of Eq. \eqn{C2} when continuous indices are replaced
by discrete N-dimensional ones  reads \cite{OlgaBeppe2}
\be
\sum_{m=1}^{N}K_l^{n m } K_m^{s k}  = \sum_{m=1}^{N}K_m^{n s }
K_l^{n k} \label{discr}
\ee
 The
associativity equation~(\ref{C2}) has a {\it hidden} symmetry,
 the scaling transform. Namely, given a solution, $K(\vec x_1, \vec x_2, \vec x_3)$, of the
equation~(\ref{C2}) the new kernel
\begin{equation}\label{C3a}
K_{\lambda}(\vec x_1,\vec x_2, \vec x_3)=\lambda K(\vec x_1,\vec
x_2,\vec x_3)
\end{equation}
is also a solution. The scaling transform of the kernel can be
induced  transforming  the quantizer and dequantizer as
\begin{eqnarray}\label{C3b}
&&\hat U(\vec x)\rightarrow\hat U_{\lambda}(\vec
x)=\lambda\hat U(\vec x), \nonumber\\
&&\\
&&\hat D(\vec x)\rightarrow\hat D_{\lambda}(\vec
x)=\lambda^{-1}\hat D(\vec x)
\end{eqnarray}
in such a way to  preserve the metrical  condition~(\ref{eq.3}).
The symmetry of Eq.~(\ref{C2}) can be described by a
transformation kernel. In fact, given a  pair $\left(\hat U(\vec
x), \hat D(\vec x)\right)$ and another pair $ \left(\hat U_1(\vec
y), \quad \hat D_1(\vec y)\right) $  which provides a new symbol
for a given operator $\hat A$
\begin{equation}\label{C4}
f_A(\vec y)=\mbox{Tr}\Big(\hat U_1(\vec y)\hat A\Big)
\end{equation}
and
\begin{equation}\label{C5}
\hat A=\int f_A(\vec y)\hat D_1(\vec y) d \vec y,
\end{equation}
it is possible to establish  the relations
\begin{equation}\label{C6}
f_A(\vec y)=\int f_A(\vec x)K_1(\vec x,\vec y)d\vec x,
\end{equation}
and
\begin{equation}\label{C7}
f_A(\vec x)=\int f_A(\vec y)K_2(\vec y,\vec x)d\vec y
\end{equation}
with
\begin{eqnarray}\label{C8}
&&K_1(\vec x,\vec y)=\mbox{Tr}\Big(\hat U_1(\vec y)\hat D(\vec
x)\Big),\nonumber\\
&&K_2(\vec y,\vec x)=\mbox{Tr}\Big(\hat D_1(\vec y)\hat U(\vec
x)\Big).
\end{eqnarray}
For the dual pair $\left(\hat U_1(\vec x)=\hat D(\vec x),\quad
\hat D_1(\vec x)=\hat U(\vec x)\right)$ the dual symbol of an
operator $\hat A$ reads
\begin{equation}\label{C9}
f_A^{(d)}(\vec x)=\mbox{Tr}\Big(\hat D(\vec x)\hat A\Big)
\end{equation}
and the operator reconstruction formula is
\begin{equation}\label{C10}
\hat A=\int\Big(\mbox{Tr}\Big(\hat D(\vec x)\hat A\Big)\Big)\hat
U(\vec x)d\vec x.
\end{equation}
Then equations~(\ref{C8}) imply  that
\begin{equation}\label{C11}
f_A^{(d)}(\vec x)=\int f_A(\vec x_1)(\mbox{Tr}(\hat D(\vec x)\hat
D(\vec x_1)))d\vec x_1.
\end{equation}
The dual kernel of Eq. ~(\ref{C3}), corresponding to the
transformed pair of quantizer and dequantizer, reflects the
symmetry of equation~(\ref{C2}). Thus, the generic
formulae~(\ref{C8}), (\ref{C9}) read,  for the transition from the
initial star-product to its dual one,
\begin{eqnarray}\label{C11a}
&&K_1(\vec x,\vec y)=\mbox{Tr}\Big(\hat D(\vec y)\hat D(\vec
x)\Big),\nonumber\\
&&K_2(\vec y,\vec x)=\mbox{Tr}\Big(\hat U(\vec y)\hat U(\vec
x)\Big).
\end{eqnarray}
It is worth noticing that a dual may be introduced for the
deformed star product as well. The kernel which is dual to the
deformed one \eqn{D7}is represented  by
\begin{equation}\label{D8}
K^{(K)}_d(\vec x_1,\vec x_2, \vec x)=\mbox{Tr}\Big(\hat U(\vec
x_1)\hat K\hat U(\vec x_2) \hat D(\vec x)\Big).
\end{equation}
Also in this case the associativity \eqn{C2} may be easily
verified.

\subsection{The dual star product and  quantum observables}
To better understand the physical meaning of the dual star-product
let us consider   the  mean value of a quantum observable $\hat A$.
In tomographic representation (see below Section 5) this was shown
to be \cite{Olga97}
\begin{equation}\label{Y1}
\langle\hat A\rangle=\mbox{Tr}\hat \rho\hat A=\int
w(X,\mu,\nu)f_A(X,\mu,\nu)d X\,d\mu\,d\nu
\end{equation}
where, comparing with standard definitions of the mean value, the
function $ f_A(X,\mu,\nu)$ can be seen to be given by
\begin{equation}\label{Y2}
f_A(X,\mu,\nu)=\mbox{Tr}\hat A\frac{1}{2\pi}\exp[i(X-\mu\hat
q-\nu\hat p)].
\end{equation}
  This is nothing but the
symbol of the observable $\hat A$ in the {\it dual} tomographic
scheme, namely,
\begin{equation}\label{Y3}
f_A(X,\mu,\nu)\equiv f^d_A (X,\mu,\nu)= \mbox{Tr}\Big(\hat
D(X,\mu,\nu)\hat A\Big),
\end{equation}
while the tomogram $ w(X,\mu,\nu)$ is the symbol of the density
matrix, in tomographic representation. Therefore, according to
\eqn{Y1}, the mean value of an observable $\hat A$ is given by the
product of the symbol of the density matrix and the symbol of the
observable in the dual representation, appropriately integrated. In
fact, it can be shown that this observation is true in general.
 For any
star-product scheme the  mean value of an observable $\hat A$ can
be always written as
\begin{equation}\label{Y4}
\langle \hat A\rangle=\mbox{Tr}(\hat\rho\hat A)=\int f_{\rho}(\vec
x)f^d_A(\vec x)\,d\vec x
\end{equation}
where $f_\rho$ is the symbol of $\hat \rho$ in a given
quantization scheme, while $f^d_A$ is the symbol of the observable
$\hat A$ in the dual scheme. Eq. \eqn{Y3} is easily verified
substituting back the definitions of $f_\rho$  and  $f^d_A$ and
using the compatibility  condition \eqn{eq.3}.  This consideration
points out the different nature   of density states and
observables. Though both are hermitian operators, they possess a
different interpretation within  the operator  algebra, the former
being linear functionals over the latter. Indeed, states are a
convex body in $ \mathcal{G}^*(U)$, the dual of the Lie algebra
$\mathcal{G}(U)$ of the unitary group $U$. The duality
transformation is again applied to the ambient space of states and
observables $\mathcal{G}^*(U)\times \mathcal{G}(U)\rightarrow
\mathcal{G}^{**}(U)\times \mathcal{G}^*(U)$. In the star product
approach, their symbols are related to dual quantum schemes.  In
self-dual schemes (Weyl-Moyal-Wigner) this difference is hidden.

\section{Class of solutions of the Jacobi identity}
It is well known that, as   any other  Lie bracket defined via an
associative product, the commutator bracket
\begin{equation}\label{J1}
[\hat A,\hat B]=\hat A\hat B-\hat B\hat A
\end{equation}
satisfies the Jacobi identity
\begin{equation}\label{J2}
\Big[[\hat A,\hat B],\hat C\Big]+\Big[[\hat B,\hat C],\hat
A\Big]+\Big[[\hat C,\hat A]\hat B\Big]=0.
\end{equation}
This in turn   gives rise to an corresponding  identity for the
symbols of the operators. To this, following~\cite{OlgaBeppe2} let
us introduce the symbol of the commutator
\begin{equation}\label{J4}
f_{[A, B]}(\vec x)=\mbox{Tr}\Big([\hat A,\hat B]\hat U(\vec
x)\Big).
\end{equation}
One has
\begin{equation}\label{J4a}
f_{[A, B]}(\vec x)=-f_{[B,A]}(\vec x).
\end{equation}
One can rewrite the symbol in the form
\begin{equation}\label{J5}
f_{[A, B]}(\vec x)=\Big(f_A\ast f_B\Big)(\vec x)-\Big(f_B\ast
f_A\Big)(\vec x).
\end{equation}
The integral form of this relation reads
\begin{equation}\label{J6}
f_{[A, B]}(\vec x)=\int K_-(\vec x_1,\vec x_2,\vec x)
f_A(x_1)f_B(\vec x_2)d \vec x_1\,d \vec x_2
\end{equation}
where
\begin{equation}\label{J8}
K_-(\vec x_1,\vec x_2,\vec x)=K(\vec x_1,\vec x_2,\vec x)-K(\vec
x_2,\vec x_1,\vec x)=\mbox{Tr}\Big([\hat D(\vec x_1),\hat D(\vec
x_2)]\hat U(\vec x)\Big).
\end{equation}
Due to Eqs. ~(\ref{J2}) the kernel~(\ref{J8}) inherits a  Jacobi
identity for the Lie bracket on the symbols
\begin{equation}\label{J9}
f_{[[A,B],C]}(\vec x)+f_{[[B,C],A]}(\vec x)+f_{[[C,A],B]}(\vec
x)=0.
\end{equation}
In terms of the kernel~(\ref{J8}) this identity reads
\begin{eqnarray}\label{J10}
&&\int \Big(K_-(\vec x_1,\vec x_2,\vec y)K_-(\vec y,\vec x_3,\vec
x_4)+ K_-(\vec x_2,\vec x_3,\vec y)K_-(\vec y,\vec x_1,\vec
x_4)\nonumber\\
 &&+ K_-(\vec x_3,\vec x_1,\vec y)K_-(\vec y,\vec
x_2,\vec x_4)\Big)d\vec y=0
\end{eqnarray}
which is the anti-symmetrized version of Eq. \eqn{C2}. When the
parameters are discrete,  this equation provides the Jacobi
identity for the structure constants of a Lie algebra. In fact,
for a $N$ dimensional Lie algebra with generators $\hat L_i$ the
commutation relations read
\begin{equation}\label{J11}
[\hat L_i,\hat L_j]=C_{i j}^l \hat L_l
\end{equation}
where the sum over repeated indices is understood. The Jacobi
identity reads here
\begin{equation}\label{J12}
[[\hat L_i,\hat L_j],\hat L_k]+[[\hat L_j,\hat L_k],\hat
L_i]+[[\hat L_k,\hat L_i],\hat L_j]=0
\end{equation}
and using ~(\ref{J11}) we get
\begin{equation}\label{J13}
C_{i j}^l C_{l k}^m\hat L_m+C_{j k}^l C_{l i}^m\hat L_m+ C_{k i}^l
C_{l j }^m \hat L_m=0.
\end{equation}
Since the generators $\hat L_m$ are a  basis in the Lie algebra,
this  implies that the structure constants satisfy the quadratic
equations (Jacobi identity)
\be
\sum_{l=1}^N C_{i j}^l C_{l k}^m+C_{j k}^l C_{l i}^m+C_{k i}^l
C_{l j}^m=0
\ee
where we have restored the summation symbol  to compare
with~(\ref{J10}). Changing notation
\begin{equation}\label{J14}
C\rightarrow K_-,\quad i\rightarrow\vec x_1, \quad j\rightarrow\vec
x_2,\quad l\rightarrow\vec y, \quad k\rightarrow \vec x_3, \quad
m\rightarrow \vec x_4
\end{equation}
and replacing the sum over $l$ with an  integral over $\vec y$ one
can see that~(\ref{J14}) is identical to~(\ref{J10}). Thus, writing
the Jacobi identity as an algebraic equation for the structure
constants of a Lie algebra, one can ask what are the solutions
(possibly all) for these equations from a novel point of view.
Namely the problem is translated into the search of compatible
pairs of quantizers $\hat D(\vec x)$ and dequantizers $\hat U(\vec
x)$; solutions of the Jacobi identity will then be given in the
form~(\ref{J8}).

We also notice that the dual formula provides other solutions of
the Jacobi identity
\begin{equation}\label{J16}
K^{(d)}_-(\vec x_1,\vec x_2,\vec x)=\mbox{Tr}\Big([[\hat U(\vec
x_1),\hat U(\vec x_2)],\hat D^\dag(\vec x)]\Big).
\end{equation}
Moreover, since any solution of the associativity equation for the
kernel of the star-product determines a solution of the Jacobi
identity for the structure constants, one can use the $K$-deformed
kernels~(\ref{D7}) and~(\ref{D8}) to obtain yet other solutions of
the Jacobi identity. In fact, the kernels given by
\begin{equation}\label{L1}
K_-^{(k)}(\vec x_1,\vec x_2,\vec x)=\mbox{Tr}\Big([\hat D(\vec
x_1),\hat D(\vec x_2)]_K \hat U^\dag(\vec x)\Big).
\end{equation}
and
\begin{equation}\label{L2}
K_-^{(k d)}(\vec x_1,\vec x_2,\vec x)=\mbox{Tr}\Big([\hat U(\vec
x_1),\hat U(\vec x_2)]_K\hat D^\dag(\vec x)\Big),
\end{equation}
where
\begin{equation}\label{L3}
[\hat D(\vec x_1),\hat D(\vec x_2)]_K=\hat D(\vec x_1)\hat K\hat
D(\vec x_2)-\hat D(\vec x_2)\hat K\hat D(\vec x_1)
\end{equation}
and
\begin{equation}\label{L4}
[\hat U(\vec x_1),\hat U(\vec x_2)]_K=\hat U(\vec x_1)\hat K\hat
U(\vec x_2)-\hat U(\vec x_2)\hat K\hat U(\vec x_1)
\end{equation}
satisfy ~(\ref{J10}) and $[~,~]_K$ is the  $K$-deformed commutator.
Of course, because this deformation does not provide us with the
most general associative algebras, we may use alternative
deformations along the lines of \cite{CGM}.

Let us note that, in finite dimensions, on using Ado's theorem on
the possibility of realizing any Lie algebra as an algebra of
matrices and using the cotangent bundle of the identified algebra
of matrices, in principle we may always realize this program. The
relevance of our proposal relies on the possibility of
constructing, by the present procedure, infinite dimensional
double Lie algebras.

\subsection{An illustrative example: the U(2) Lie algebra}
To see our previous constructions  at work let us consider  a well
known example, the  $U(2)$ Lie algebra. Proceeding backwards from
the Lie algebra we want to  show how the corresponding quantizer
and dequantizer look like. Then, exchanging their role as in Eq.
\eqn{J16} we will find a new Lie algebra which is dual to the
starting one. These being compatible as Lie algebras, their sum is
still a Lie algebra, known as the double (see for example
\cite{AGMM}.

 Let
us consider the  structure constants of the group U(2)
\begin{equation}\label{R1}
C_{i j}^k=
  \begin{cases}
   \varepsilon_{i jk }, & \text{  i,\,j,\,k = 1,\,2,\,3} \\
   0 & \text{otherwise}.
  \end{cases}
\end{equation}
The generators of  $U(2)$  may be given in terms of Pauli matrices
plus the $2\times 2$ identity matrix
\beqa\label{R6}
S_j&=&\frac{\sigma_j}{2}, ~~~j\ne 0\nn\\
S_0=\sigma_0
\eeqa with
\begin{equation}\label{R2}
\sigma_0=
\begin{pmatrix}
  1&0 \\
  0&1
\end{pmatrix},\quad\sigma_1=
\begin{pmatrix}0&1\\ 1&0
\end{pmatrix}, \quad\sigma_2=\begin{pmatrix}0&-i\\
i&0 \end{pmatrix},\quad\sigma_3=\begin{pmatrix} 1&0\\
0&-1\end{pmatrix}
\end{equation}
We recall their commutation rules
\begin{equation}\label{R3}
[\sigma_0,\sigma_j]=0;\quad [\sigma_j,\sigma_k]=2i\varepsilon_{j k
m}\sigma_{m};
\end{equation}
while the associative product obeys
\begin{equation}\label{R4}
\sigma _i \sigma _0 = \sigma_0\sigma_i=
\sigma_i;~~~\sigma_j\sigma_k=\delta_{j k}\sigma_0 +i\varepsilon_{j
k m}\sigma_m,\quad j,\,k,\,m=1,\,2,\,3 ,
\end{equation}
(here we sum over the index $m$). As for the traces
\begin{equation}\label{R5}
\mbox{Tr}\sigma_0=2,\quad \mbox{Tr}\sigma_j=0.
\end{equation}
Thus one has
\begin{equation}\label{R7}
[S_j,S_k]=i\varepsilon_{j k m}S_m.
\end{equation}
Now we select a basis in the vector space underlying
$\mathcal{U}(2)$ and a basis for its dual vector space. This
selection may be operated choosing a non-degenerate (0,2)-tensor
on $\mathcal{U}(2)$, which realizes the pairing between the two
vector spaces.
 If the pairing between
$\mathcal{U}(2)$ and its dual is provided by the twice the trace,
\be
<~,~>=2 \Tr
\ee
the dequantizer and the quantizer may be
seen to be respectively
\begin{equation}\label{R8}
\hat U(x)=\{\hat U(0)=\half\sigma_0,\quad \hat U(1)=\half\sigma_1,
\quad\hat U(2)=\half\sigma_2, \quad\hat U(3)=\half\sigma_3\}
\end{equation}
and
\begin{equation}\label{R9}
\hat D(x)=\hat U(x)
\end{equation}
where we have introduced the notation $x=(0,1,2,3)$. Indeed, they
satisfy the compatibility condition \eqn{eq.3}
\begin{equation}\label{R10}
<(\hat U(j),0), (0,\hat D(k))>= 2 \mbox{Tr}\Big(\hat U^\dag
(j)\hat D(k)\Big)=\frac{1}{2}\mbox{Tr}
\sigma_j\sigma_k=\delta_{jk},
\end{equation}
where $\delta_{jk}$ is the Kroneker delta-function and reproduce
the correct structure constants by means of the kernel
\begin{equation}\label{R14}
K_-(j,k,m)=\frac{1}{4}\mbox{Tr}([\sigma_j,\sigma_k]\sigma_m);\quad
(j,\,k,\,m=0,\,1,\,2,\,3).
\end{equation}
 The symbol of any operator represented by a matrix A reads
\begin{equation}\label{R11}
f_A(x)=2~ \mbox{Tr}(\hat U(x) A)
\end{equation}
 Thus one has
\begin{equation}\label{R12}
f_A(0)=\mbox{Tr}A,\quad f_A(1)=\mbox{Tr}A\sigma_1,\quad
f_A(2)=\mbox{Tr}A\sigma_2,\quad f_A(3)=\mbox{Tr}A\sigma_3.
\end{equation}
while the star product is given by the kernel \eqn{eq.7} with
$\hat U(x)$ and $\hat D(x)$ given by~(\ref{R8}),~(\ref{R9}).

What about other choices of the dual basis? The dual vector space,
when the pairing is given by the trace, is endowed with a Lie
algebra structure which is again $ \mathcal{U}(2)$, namely, with
our choices, $\mathcal{U}(2)$ is selfdual. Indeed, having chosen
 quantizer and dequantizer to coincide,  the dual
antisymmetric kernel may be seen to be
\begin{equation}\label{R15}
K^{d}_-(j,k,m)=K_-(j,k,m)=\frac{1}{4}\mbox{Tr}\Big([\sigma_j,\sigma_k]\sigma_m\Big)
\end{equation}
which is (in view of~(\ref{R7}))
\be \label{R16}
K^{d}_-(j,k,m)=
  \begin{cases}
    \varepsilon_{j k m} & \text{ j,\,k,\,m=1,\,2,\,3} \\
    0 & \text{otherwise}.
  \end{cases}
\ee
 Thus we get
back the  $U(2)$ structure constants.

In $V^*\times V$ we may  choose a different dual basis. For
instance we may choose dual bases like
\beqa
\hat D(j)&=&-\frac{i}{2} \sigma_j\\
\hat U(j)&=&\frac{1}{2} \left(\sigma_j + i \varepsilon_{j k
3}\sigma_k \right)
\eeqa
with the pairing given by
\be
<(\hat U(j),0), (0,\hat D(k))>= 2 \mbox{Im~Tr}\Big(\hat U^\dag
(j)\hat D^\dag(k)\Big)=\delta_{jk}
\ee
 Indeed, the antisymmetric kernel
\be
K_-(ijk)= 2 \rm{Im}  \Tr [\hat D(i),\hat D(j)] \hat U^\dag(k)
\ee
 reproduces the $SU(2)$ structure
constants.  Its dual may be easily calculated yielding
\be
K^{d}_-(ijk)= 2 \rm{Im}  \Tr[\hat U(i),\hat U(j)] \hat
D^\dag(k)=\varepsilon_{ijl}\varepsilon^{lk3}.
\ee
These can be recognized to be the structure constants of the
$SB(2,\mathbb{C})$ Lie algebra, namely the algebra of the $2\times
2$ upper triangular complex matrices. The resulting algebra
$V^*\oplus V= \mathcal{SB}(2,\mathbb{C})\oplus \mathcal{SU}(2)$  is
isomorphic to the metrical double Lie algebra
$\mathcal{SL}(2,\mathbb{C}) $ with Manin decomposition (see for
example\cite{mss}).

\subsection{K-deformed products: examples}
At the beginning of the section we have argued  that also the
deformed kernel \eqn{L1} satisfies an associativity condition
which provides  the Jacobi identity for the structure constants of
a Lie algebra. Therefore deformations  may be put to work to
generate different Lie algebras starting from a given one.

In this section we give an explicit example of this statement
showing that, starting from the $SO(3)$ Lie algebra we recover all
unimodular   3-d Lie algebras, along the lines previously illustated
in  \cite{marpergrab} (for a review see \cite{GLMV02},
\cite{MMVZ94}).

There, the Lie algebras under consideration  where realized in terms
of Poisson brackets defined on their dual. More precisely it was
observed that any  real finite-dimensional Lie algebra ${\mathcal
G}$ with Lie bracket $[{\cdot},{\cdot}]$ defines in a natural way a
Poisson structure $\{{\cdot},{\cdot}\}$, on the dual space
${\mathcal G}^*$ of ${\mathcal G}$. One is allowed to think of
${\mathcal G}$ as a subset of linear functions within the ring of
smooth functions $C^\infty ({\mathcal G}^*)$. Choosing a linear
basis $\{E_i\}_1^n$ of ${\mathcal G}$, and identifying them with
linear coordinate functions $x_i$ on ${\mathcal G}^*$ by means of
$x_i(x) = <x,E_i>$ for all $x\in {\mathcal G}^*$, we define the
fundamental brackets on ${\mathcal G}^*$ by the expression
$\{x_i,x_j\}_{{\mathcal G}}=c_{ij}^k x_k$ where $[E_i,E_j]=c_{ij}^k
E_k$ and $c_{ij}^k$ denote the structure constants of the Lie
algebra. On this basis the equivalence classes of all three
dimensional Lie algebras are seen to be characterized by the Casimir
form
\be
\alpha = h(x_1\,dx_2-x_2\,dx_1) +  \half\, d
(ax_1^2+bx_2^2+cx_3^2) \label{alph}
\ee
by means of
\be
\{x_j, x_k \}= \varepsilon_{jkl}\alpha \left(\frac{\del}{\del
x^l}\right)
\ee
 with the real parameters $h,a,b,c$ appropriately
selected. This yields the Poisson brackets
\be
\{x_1,x_2\}=cx_3,~~~~ \{x_2,x_3\}=ax_1 -hx_2,~~~~
\{x_3,x_1\}=bx_2+hx_1 \label{classbr}
\ee
with the Jacobi identity encoded by
\be
d\alpha\wedge \alpha = 2h\,c\,x_3\;dx_1\wedge dx_2\wedge dx_3 = 0
\ee
which in turn holds true if and only if $hc=0$. Thus we have two
essentially different classes of algebras: those corresponding to
a closed Casimir form ($h=0$, case A), and ($c=0$, case B), those
corresponding to a  Casimir form which is not closed.

In case A the parameters $a,b,c$, when different from zero, may be
all normalized to modulus one. This grouping includes six
different isomorphism classes of Lie algebras:

\begin{itemize}
\item[A.1] $su(2)\simeq so(3)$ with $a,b,c$ all different from 0 and of the
same sign. A basis can be chosen so that $a=b=c=1$.
\item[A.2] $e(2)$, the algebra of the Euclidean group in two dimensions, which
may be obtained by contraction from the previous class, say $a\to
0$.
\item[A.3] $sl(2,\real)\simeq su(1,1)\simeq so(2,1)$, with $a,b,c$ all
different from 0 and one of them of different sign.
\item[A.4] $iso(1,1)$, the Poincar\'{e} algebra in two dimensions, which may be
obtained by contraction from the previous algebra.
\item[A.5] $h(1)$, the Heisenberg-Weyl algebra, with only one parameter
different from zero, for example $c> 0$. It may be obtained by
further contraction from both $e(2)$ and $iso(1,1)$. \item[A.6]
The abelian algebra with a=b=c=0.
\end{itemize}
All these brackets may be written in the form
\be
\{x_i,x_j\}= \varepsilon_{ijl}\frac{\del \mathcal{C} }{\del x_l}=
\varepsilon_{ijl} d \mathcal{C} \left( \frac{\del}{\del
x_l}\right)\label{pa}
\ee
where
\be
\alpha=d \mathcal{C};~~~ \mathcal{C}= \frac{1}{2}(ax_1^2+b x_2^2+c
x_3^2)\label{cas}
\ee
 The second case, with $\alpha$ not exact, includes four
families of Lie algebras. We recall that all algebras of type B
have $c=0$ because of the Jacobi identity. One has
\begin{itemize}
\item[B.1] $h=1,a=b=0$, that is $sb(2,\complex)$, the Lie algebra of the
group of $2{\times}2$ upper(lower) triangular complex matrices
with unit determinant.
\item[B.2] $h=1,a=0,b=1$.
\item[B.3] $h\ne0,a=1,b=-1$.
\item[B.4] $h\ne0,a=b=1$.
\end{itemize}
Cases B.3 and B.4 are one-parameter families, as it is impossible to
put all parameters equal to one with a similarity transformation. We
will refer collectively to the previous four classes of algebras as
${\mathcal G}_h$.

The K-deformation procedure provides a tool to derive all the
three dimensional Lie algebras above. In fact, as for the $SU(2)$
Lie algebra  the pair $\hat U(x), \hat D(x)$ which generate the
$SO(3)$ Lie algebra can be chosen to be
\be
\hat U(x) \equiv U(j)=L_j, ~~~ \hat D(x)\equiv D(j) = \half U(j)
\ee
Thus the general scheme of K deformed structure constants can be
applied.
 Let us consider  the generators of $SO(3)$
\be
L_1=\begin{pmatrix}
  0 & 0 & 0 \\
  0 & 0 & -1 \\
  0 & 1 & 0
\end{pmatrix}
,~~ L_2=\begin{pmatrix}
  0 & 0 & 1 \\
  0 & 0 & 0 \\
  -1 & 0 & 0
\end{pmatrix}
,~~ L_3=
\begin{pmatrix}
  0 & -1 & 0 \\
  1 & 0 & 0 \\
  0 & 0 & 0
\end{pmatrix}
\ee
and a $3\times 3$ symmetric  matrix $K$
\begin{equation}\label{R17}
K=\begin{pmatrix} \lambda_1 &\mu_1& \mu_3\\
\mu_1&\lambda_2&\mu_2\\
\mu_3&\mu_2&\lambda_3
\end{pmatrix}
\end{equation}
Using the $K$ deformed Lie bracket $[L_i,L_j]_K\equiv L_i K L_j -
L_j KL_i $ we obtain the algebra
\beqa
{[L_1,L_2]}_K &=& \mu_3 L_1 +  \mu_2 L_2+ \lambda_3 L_3\nn\\
{[L_2,L_3]}_K &=& \lambda_1 L_1 +  \mu_1 L_2+ \mu_3 L_3 \nn\\
{[L_3,L_1]}_K &=& \mu_1 L_1 +  \lambda_2 L_2+ \mu_2 L_3 \label{la}
\eeqa
It can be verified that all type A algebras are obtained by suitable
choices of the parameters. Indeed, the Lie algebra represented by
\eqn{la} may be written in terms of Poisson brackets by setting
\be
d\mathcal{C}_K= \begin{pmatrix}
  dx_{1} & dx_{2} & dx_{3}
\end{pmatrix}
\begin{pmatrix}
  \lambda_{1} & \mu_{1} & \mu_{3} \\
  \mu_{1} & \lambda_{2} & \mu_{2} \\
  \mu_{3} & \mu_{2} & \lambda_{3}
\end{pmatrix}
\begin{pmatrix}
  x_{1} \\
  x_{2} \\
  x_{3}
\end{pmatrix}
\ee
Then $\{x_i,x_j\}= \varepsilon_{ijl} d \mathcal{C}_K\bigl(
\frac{\del}{\del x_l}\bigr)$. On using a diagonalizing
transformation for $K$ we may reduce $d\mathcal{C}_K$ to the form
\eqn{cas}. This analysis shows that the quadratic form associated
with $K$ is a Casimir for the Lie algebra we are going to define.

The algebras of type B do not posses a Casimir function, but the
procedure to obtain all of them via a $K$ deformation may still be
applied. Of course, we have to start from a type B algebra to
generate all the others. We consider the Lie algebra  B.4 of our
classification.
The generators may be chosen in the form
\be
X_1=\begin{pmatrix}
  0 & 0 & 1 \\
  0 & 0 & 0 \\
  0 & 0 & 0
\end{pmatrix}
,~~ X_2=\begin{pmatrix}
  0 & 1 & 0 \\
  0 & 0 & 0 \\
  0 & 0 & 0
\end{pmatrix}
,~~ X_3=
\begin{pmatrix}
  h & 0 & 0 \\
  0 & 0 & 1 \\
  0 & -1 & 0
\end{pmatrix}
\ee
 The other Lie algebras
may be obtained, as previously, $K$-deforming  the commutator with
a suitable $K$ matrix. Imposing Jacobi identity this may be
checked to be of the form
\be
K=\begin{pmatrix}
  \alpha & \beta & \gamma \\
  0 & \epsilon & \phi \\
  0 & \zeta & \iota
\end{pmatrix} .
\ee
All  type B algebras are thus reproduced with an appropriate
choice of the parameters. For example, the type B.1 algebra
$\mathcal{SB}(2,C)$ is obtained  with $\epsilon=\iota=0$ and
$h\alpha=1-\phi=1+\zeta$.

However, our duality approach, via the formula \eqn{L2}, opens up
other possibilities.

To provide an example in infinite dimensions we consider a
deformation of the Weyl-product. Using the pair $\hat U(q,p)$ and
$\hat D(q,p)$ for the Weyl product as in \cite{Patr} one can
define the structure constants of an infinite-dimensional Lie
algebra  as
\begin{eqnarray}\label{R19}
& &
K_-(q_1,p_1,q_2,p_2,q_3,p_3)=\frac{2}{\pi}\mbox{Tr}\Bigl[\Bigl(\hat{
\mathcal{D}}(2\alpha_1)f(a^+ a)\hat{\mathcal{D}}(-2\alpha_2) \nn \\
 && -\hat{\mathcal{D}}(2\alpha_2)f(a^+a)\hat{ \mathcal{D}}(-2\alpha_1)
 \Bigr)\hat{
\mathcal{D}}(2\alpha_3)(-1)^{a^+a}\Bigr],
\end{eqnarray}
where
$$\alpha_m=\frac{q_m+i p_m}{\sqrt2},\quad m=1,2,3.$$
Here $f(a^+a)$ plays the role of the deformation operator $\hat
K$, $\hat D(\alpha)$ is the Weyl system operator. Thus  we find a
deformation using the formalism of nonlinear $f$-deformed
oscillators ~\cite{PhysScr}. We shall elaborate on the duality
emerging from this example elsewhere.

\section{The tomographic setting and its classical limit}
With respect to quasi-distributions on phase-space, tomograms have
the property of being positive both for classical and quantum
systems, therefore they furnish a convenient setting to analyze
the quantum-classical transition, because we deal with the same
kind of objects.
 Let us review the construction of the symplectic
tomographic map using explicitly the Planck's constant $\hbar$ as
deformation parameter. In this case we choose $\vec
x=(x_1,x_2,x_3)$ with $x_1=X,\,x_2=\mu,\,x_3=\nu\,\in\,R$. The
dequantizer is taken in the form
\begin{equation}\label{Q1}
\hat U(\vec x):=\hat U(X,\mu,\nu)=\delta(X-\mu \hat q-\nu \hat p
).
\end{equation}
Here $\hat q$ and $\hat p$ are position and momentum operators.
The random variable $X$ describes the position of a particle in a
reference frame in its phase-space. But this reference frame is
squeezed (parameter s) and rotated (angle $\theta$), so that
\begin{equation}\label{Q2}
\mu=s\cos\theta,\,\nu=s^{-1}\sin\theta.
\end{equation}
The quantizer operator is given by
\begin{equation}\label{Q3}
\hat D(\vec x)=\hat D(X,\mu,\nu)=\frac{\hbar}{2\pi}\exp
[i(X-\mu\hat q-\nu\hat p)].
\end{equation}
One can check that this choice of  $\hat U(\vec x),~\hat D(\vec
x)$ fulfills the compatibility condition \eqn{eq.3}
\begin{equation}\label{Q4}
\mbox{Tr}\left(\hat U(X,\mu,\nu)\hat
D(X^\prime,\mu^\prime,\nu^\prime)\right)=\delta(X-X^\prime)\delta(\mu-\mu^\prime)\delta(\nu-\nu^\prime).
\end{equation}
Thus,   to  operators $\hat A$ in the Hilbert space $\mathcal{H}$
one can  associate  the function (tomographic symbol or tomogram)
\begin{equation}\label{Q5}
w_A(X,\mu,\nu)=\mbox{Tr}\hat A\delta(X-\mu\hat q-\nu\hat p).
\end{equation}
Conversely, the tomogram allows to reconstruct  the operator via
the relation
\begin{equation}
\hat A=\frac{\hbar}{2\pi}\int w_A(X,\mu,\nu)\exp[i(X-\mu\hat q-\nu
\hat p)]dX\,d\mu\,d\nu. \label{tomsymb}
\end{equation}
The tomogram of a pure state with wave function $\psi(x)$ can be
calculated as in \eqn{tomsymb}, with the operator $\hat A$ replaced
by the density operator $\hat \rho= (|\psi><\psi|)/<\psi|\psi>$
\begin{equation}\label{Q10}
w_\psi(X,\mu,\nu)=\frac{1}{2\pi\hbar|\nu|}\left|\int\psi(y)\exp(\frac{i\mu
y^2}{2\nu \hbar}-\frac{iX y}{\hbar\nu})dy\right|^2.
\end{equation}
 According to \eqn{eq.6} the star-product of two tomograms is
\be
w_A\ast w_B(x)=\int w_A(x_1) w_B(x_2) K(x_1,x_2,x) \hbar^2 dx_1 dx_2
\ee
where we have introduce the collective variables $x=(X,\mu,\nu)$,
$x_i=(X_i,\mu_i,\nu_i)$ and, following Eq. \eqn{eq.7}, the kernel is
  given by
\begin{equation}\label{Q7}
K(x_1,x_2,x)=\mbox{Tr}\left(\hat D(x_1)\hat D(x_2,)\hat U^\dag(x)
\right),
\end{equation}
which explicitly reads
\begin{eqnarray}\label{Q8}
&&K(x_1,x_2,x)=\mbox{Tr}
\Big(\frac{\hbar^2}{4\pi^2}\exp(iX_1+iX_2-i\mu_1\hat q_1+i\nu_1\hat p_1)\nonumber\\
&&\times\exp(-i\mu_2\hat q_2-i \nu_2\hat
p_2)\delta(X-\mu\hat q-\nu\hat p) \Big)\nonumber\\
&&= \frac{\hbar^2}{4\pi^2}\delta\Big(
\nu(\mu_1+\mu_2)-\mu(\nu_1+\nu_2)\Big)
\times\exp\Big(iX_1+iX_2+\frac{i\hbar}{2}(\nu_1\mu_2-\nu_2\mu_1)-
i\frac{(\nu_1+\nu_2)X}{\nu}\Big). \nn\\
\end{eqnarray}
Notice that we have included the factor  $\hbar^2$ in the measure,
not in the kernel, in such a way to obtain   a kernel with the same
dimensions of its classical analogue, as we will see below. The
kernel is the product of two factors. The first one
\begin{eqnarray}\label{Q11}
&&K_{\rm cl}(x_1,x_2,x_3)= \frac{1}{4\pi^2}\delta\Big(
\nu_3(\mu_1+\mu_2)-\mu_3(\nu_1+\nu_2)\Big)
\nonumber\\
&&\times\exp\Big(iX_1+iX_2-i\frac{(\nu_1+\nu_2)X_3}{\nu_3}\Big)
\end{eqnarray}
 furnishes  the point-wise product
of functions in phase space as we will see explicitly in a moment.
The second factor
\be
f(\hbar)=  \exp[\frac{i\hbar}{2} (\nu_1 \mu_2 - \nu_2 \mu_1)].
\label{fh}
\ee
is  antisymmetric and depends explicitly on the Planck constant.
This factorization shows very clearly that quantum mechanics
modifies the structure functions of the classical associative
product by means of a factor proportional to the exponential of a
symplectic area.
\subsection{The classical product in tomographic
representation}
Given two functions $A(q,p)$ and $B(q,p)$ on the
classical phase space  their  pointwise product has the integral
representation
\be
A(p,q) \cdot B(p,q)= \int A(p_1,q_1)B(p_2,q_2) \delta( q-q_1)
\delta( q-q_2) \delta( p-p_1) \delta( p-p_2) dq_1 dq_2 dp_1 dp_2
\label{p1}
\ee
where the kernel  reads
\be K(q_1, p_1, q_2, p_2, q,p) =\delta(
q-q_1) \delta( q-q_2) \delta( p-p_1) \delta( p-p_2).\label{p2}
\ee
This is symmetric with respect to permutations $1 \leftrightarrow
2$. What  is the kernel of this product in the  tomographic
representation? To this, let us  introduce the  tomographic
symbols  of the functions $A(q,p)$ and $B(q,p)$. They are
respectively given by the Radon transform
\beqa
w_A(X,\mu,\nu)&=& \int A(q,p) \delta(X-\mu q - \nu p) \frac{dq
dp}{2\pi}\nn\\
w_B(X,\mu,\nu)&=& \int B(q,p) \delta(X-\mu q - \nu p) \frac{dq
dp}{2\pi}. \label{P3}
\eeqa
This  is  invertible with inverse
\beqa
A(q,p)&=& \frac{1}{2\pi}\int w_A(X,\mu,\nu) \exp[i(X-\mu q - \nu
p)] dX
d\mu d\nu \nn\\
B(q,p)&=& \frac{1}{2\pi}\int w_B(X,\mu,\nu) \exp[i(X-\mu q - \nu
p)] dX d\mu d\nu. \label{P4}
\eeqa
Using \eqn{P3} the classical tomogram associated to the product
$A\cdot B$ reads
\be
w_{A\cdot B} =  \int A(q,p)\cdot B(q,p) \delta(X-\mu q - \nu p)
\frac{dq dp}{2\pi}. \ee Then, on using \eqn{p1}, \eqn{P4},  it can
be easily checked that the commutative pointwise product Eq.
\eqn{p1}  induces for the tomograms a nonlocal commutative product
whose kernel is exactly Eq. \eqn{Q11}, as announced.

\subsection{Poisson brackets in the tomographic representation}
As we have seen, quantum mechanical corrections to the pointwise
product are contained in the twist factor of the star product. From
Eq. \eqn{Q8}, \eqn{fh}, the kernel of the pointwise product in
tomographic representation acquires, up to the first order in
$\hbar$, a correcting factor. It reads \be\label{Q12}
\prod(x_1,x_2,x_3)= \frac{i\hbar}{2}(\nu_1\mu_2-\nu_2\mu_1) K_{\rm
cl}(X_1,\mu_1,\nu_1,X_2,\mu_2,\nu_2,X_3,\mu_3,\nu_3) \ee with
$K_{\rm cl}$ given by \eqn{Q11}. Thus,  the kernel corresponding to
the classical Poisson bracket in tomographic representation is given
by
\begin{eqnarray}\label{Q13}
&&P(x_1,x_2,x_3) = \lim_{\hbar \rightarrow 0} \frac{1}{i \hbar}
[K(x_1,x_2,x_3) - K(x_2, x_1, x_3)]\nn\\ &&= (\nu_1\mu_2-\nu_2\mu_1)
\frac{1}{4\pi^2}\delta\Big(
\nu_3(\mu_1+\mu_2)-\mu_3(\nu_1+\nu_2)\Big)
\exp\Big(iX_1+iX_2-i\frac{(\nu_1+\nu_2)X_3}{\nu_3}\Big).
\end{eqnarray}

In order to convince ourselves that \eqn{Q13} really represents
the canonical Poisson bracket on the phase space $\mathbb{R}^{2n}$
we may derive the Poisson bracket in tomographic representation
directly from the definition and then compare with \eqn{Q13}.

>From the inverse Radon transform \eqn{P4} we have
\beqa
\frac{\del A(q,p)}{\del q} &=& \frac{1}{2\pi}\int w_A(X,\mu,\nu)
(-i \mu) \exp[i(X-\mu q - \nu p)] dX d\mu d\nu\nn\\
\frac{\del A(q,p)}{\del p} &=& \frac{1}{2\pi}\int w_A(X,\mu,\nu)
(-i \nu) \exp[i(X-\mu q - \nu p)] dX d\mu d\nu \label{derivatives}
\eeqa
The tomogram associated to the Poisson bracket
\be
\{A(q,p), B(q,p)\}=\frac{\del A(q,p)}{\del q}\frac{\del
B(q,p)}{\del p}- \frac{\del A(q,p)}{\del p} \frac{\del
B(q,p)}{\del q}
\ee
reads
\be
w_{\{A,B\}}= \int \{A(q,p), B(q,p)\} \delta (X-\mu q - \nu p)
\frac{dq dp}{2\pi}
\ee
Therefore, upon substituting the inverse Radon transform of
derivatives \eqn{derivatives}, it may be easily checked that
\be
w_{\{A,B\}} (x)= \int w_A (x_1)  w_A (x_2) P(x_1,x_2,x) dx_1~ dx_2
\ee
where the kernel $P$ coincides with our previous result, \eqn{Q13}
and we recall that $dx_i\equiv dX_i ~d\mu_i ~d\nu_i$.

\subsection{The  Fourier representation}
We have just seen that the tomographic star product, when regarded
from the point of view of its integral kernel, only differs from the
classical one by a twist factor proportional to a symplectic area.
Here we want to show that this is not a peculiarity of the
tomographic product, but the same happens for the Moyal product.

 To this, let us  derive the star product kernel for the Moyal product
in Fourier representation. The Moyal product of two Weyl symbols
is given by the well known asymptotic formula
 \begin{equation}\label{F1}
A(q,p)\ast B(q,p)=
A(q,p)\exp\Big[i\Big(\frac{\partial}{\partial\overleftarrow{q}}\frac{\partial}{\partial
\overrightarrow{p}}-\frac{\partial}{\partial\overleftarrow{p}}\frac{\partial}{\partial
\overrightarrow{q}}\Big)\Big]B(q,p)
\end{equation}
where $\hbar=1$. Derivatives with a left arrow are understood to
act on the left while those with a right arrow  act on the right,
therefore we have
\begin{eqnarray}\label{F1a}
A\ast B &=&\exp\Big[i\Big(\frac{\partial}{\partial q_1}
\frac{\partial}{\partial p_2} -\frac{\partial}{\partial p_1}
\frac{\partial}{\partial q_2}\Big)\Big]A(q_1,p_1)B(q_2,p_2)_{
\left|{\begin{array}{c}q\equiv q_1=q_2\\p\equiv
p_1=p_2\end{array}}\right.}\nonumber\\
&=&\Big[A(q_1,p_1)B(q_2,p_2)+i\Big(\frac{\partial
A(q_1,p_1)}{\partial q_1}\frac{\partial B(q_2,p_2)}{\partial
p_2}-\frac{\partial A(q_1,p_1)}{\partial p_1}\frac{\partial
B(q_2,p_2)}{\partial q_2}\Big)\nonumber\\
&&-\frac{1}{2!}\Big(\frac{\partial}{\partial q_1}
\frac{\partial}{\partial p_2} -\frac{\partial}{\partial p_1}
\frac{\partial}{\partial q_2}\Big)^2 A(q_1,p_1)B(q_2,p_2)+\cdots
\Big]_{\left|{\begin{array}{c}q\equiv q_1=q_2\\p\equiv
p_1=p_2\end{array}}\right.}.\nonumber\\
&&
\end{eqnarray}
This may  be rewritten in the form
\begin{equation}\label{F1b}
A(q,p)\ast B(q,p)=\int A(q_1,p_1)B(q_2,p_2)K(q_1,p_1,q_2,p_2,q,p)\,
d q_1\, d q_2\, d p_1\, d p_2.
\end{equation}
where $K(q_1,p_1,q_2,p_2,q,p)$ is the Gr\"onewold
kernel~\cite{Groen}
\begin{equation}\label{F1c}
K(q_1,p_1,q_2,p_2,q,p)=\frac{1}{\pi^2}\exp[2i(p_2q_1-p_1q_2+p
q_2-p_2q+p_1q-p q_1)].
\end{equation}
The argument of the exponential can be seen to coincide with the
symplectic area of the triangle with vertices $(p,q),(p_1,q_1),
(p_2,q_2)$, if rewritten as $(p_1-p)(q_2-q) - (q_1-q)(p_2-p)$.  Now,
let us consider the star-product~(\ref{F1})  in Fourier
representation. This form corresponds to definition of Weyl symbol
in terms of the operator matrix $\tilde A (x, x')$ in position
representation
\be
A(q,p)= \int \tilde A (x=q+ \frac{u}{2}, x'= q-\frac{u}{2}) e^{-i
pu} du
\ee
and $B(q,p)$ in terms of the matrix $\tilde B (x, x')$ in position
representation
\be
B(q,p)= \int \tilde B (x=q+ \frac{u}{2}, x'= q-\frac{u}{2}) e^{-i
pu} du .
\ee
The product of matrices $\tilde A , ~ \tilde B$ is given by
\be
( \tilde A ~ \tilde B) (x, y)= \int \tilde A (x, x')  \tilde B
(x', y) dx'.
\ee
 The star-product~(\ref{F1}) of two
Weyl symbols in Fourier representation can be written in the form
\begin{equation}\label{F4}
A(\mu,\nu)\ast B(\mu,\nu)=\int
A(\mu_1,\nu_1)B(\mu_2,\nu_2)K(\mu_1,\nu_1,\mu_2,\nu_2,\mu,\nu)\, d
\mu_1\, d \mu_2\, d \nu_1\, d \nu_2,
\end{equation}
with the kernel equal to
\begin{eqnarray}\label{F5}
K(\mu_1,\nu_1,\mu_2,\nu_2,\mu,\nu)&=&\frac{1}{2\pi}\exp\Big[\frac{i}{2}
(\nu_1 \mu_2-
\nu_2 \mu_1) \Big]\nonumber\\
&&\times\delta(\mu-\mu_1-\mu_2)\delta(\nu-\nu_1-\nu_2).
\end{eqnarray}
This is the product of a twist factor represented by the exponential
of the symplectic area and   delta functions which, as we will point
out in a moment,  correspond to the pointwise product contribution.
The kernel contains an antisymmetric and a symmetric part with
respect to permutation $1\leftrightarrow 2$. The antisymmetric term
is determined by the exponent of the symplectic area in the
$\mu-\nu$ plane. So the whole kernel has the form of twisted star
product.

 As
an example we consider the Weyl symbol of the unity operator
\begin{equation}
 A_1(q,p)=1,
\end{equation}
and calculate the star product $A_1(\mu,\nu) \ast A_1(\mu,\nu)$
with
\begin{equation}\label{F6}
A_1(\mu,\nu)=\frac{1}{2\pi}\int\exp\Big[-i(\mu q+\nu p)\Big]d q\,d
p =2\pi\delta(\mu)\delta(\nu)
\end{equation}
the  Fourier component of $1$. From Eqs. ~(\ref{F4}), ~(\ref{F5})
we obtain
\begin{eqnarray}\label{F7}
A_1(\mu,\nu) \ast
A_1(\mu,\nu)&=&(2\pi)^2\int\delta(\mu_1)\delta(\nu_1)\delta(\mu_2)
\delta(\nu_2)\frac{1}{2\pi}
\exp\Big[\frac{i}{2}(\nu_1\mu_2-\nu_2\mu_1) \Big]\nonumber\\
&&\times\delta(\mu-\mu_1-\mu_2)\delta(\nu-\nu_1-\nu_2)d\mu_1\,d\mu_2\,d
\nu_1\,d\nu_2=2\pi\delta(\mu)\delta(\nu)= A_1(\mu,\nu).\nn\\
\end{eqnarray}

As for the pointwise commutative product \eqn{p1} its Fourier
representation is easily seen to be given by a nonlocal commutative
product with  kernel
 \be
K_F(\mu_1,\nu_1, \mu_2,\nu_2, \mu,\nu) =  \frac{1}{2\pi}
\delta(\mu-\mu_1-\mu_2) \delta(\nu-\nu_1-\nu_2) \label{p5}
\ee
once we consider the  Fourier
 transforms of the functions $A(q,p)$ and $B(q,p)$
\begin{equation}\label{F2}
A(\mu,\nu)=\frac{1}{2\pi}\int A(q,p)\exp[-i(\mu q+\nu p)] d q\,d
p,
\end{equation}
\begin{equation}\label{F3}
B(\mu,\nu)=\frac{1}{2\pi}\int B(q,p)\exp[-i(\mu q+\nu p)] d q\,d p.
\end{equation}
Again we find that, by working on phase-space, the classical
point-wise product and the quantum noncommutative product, when seen
through a kernel function, show  their difference in the exponential
of the symplectic area. We trust that these aspects will turn out to
be helpful when we try to generalize the construction to general Lie
groups replacing the Abelian vector groups we are using here.

Previous considerations should make clear that we have paved the way
to deal with double Lie algebras at the quantum level. These aspects
shall be taken up elsewhere.

\section{Conclusions}
We summarize the main results of our work. Deforming the associative
product on the space of matrices and using the duality symmetry we
found a class of solutions for the associativity equation of the
product kernel in factorized form (e.g. formulae \eqn{J16},
\eqn{L1}) . We then used this solution to find  a class of solutions
for the Lie algebra Jacobi identity.

We exploited the duality symmetry for the case of the tomographic
star product and we suggested a physical  interpretation for the
dual.

 A relevant aspect of this paper  is the
definition of a quantum Poisson bracket on tomograms along with its
classical limit. This is  achieved observing that with any Wigner
function we can associate a tomogram in an invertible way. The
product on Wigner functions is associated with a product on
tomograms, therefore it  induces a skew-symmetric bracket. In the
classical limit the Moyal bracket gives rise to a Poisson bracket on
phase-space.  In the analogue classical limit on tomograms this
kernel does not have the form of a bidifferential operator on $
\mathbb{R}^3$.

It would be interesting to clarify all structure constants that
are solutions of Eq. \eqn{C2} and can be obtained by the
factorization formula \eqn{eq.7}. We are presently working on this
subject.

\section*{Acknowledgement}
O.V.M. is grateful to the Russian Foundation for Basic Research
for partial support under Project No.~03-02-16408. O.V.M. and
V.I.M. thank the University ``Federico  II" of Naples for kind
hospitality.

\end{document}